\documentclass[twoside,twocolumn,9pt]{article}
\usepackage{extsizes}
\usepackage[super,sort&compress,comma]{natbib} 
\usepackage[version=3]{mhchem}
\usepackage[left=1.5cm, right=1.5cm, top=1.785cm, bottom=2.0cm]{geometry}
\usepackage{balance}
\usepackage{mathptmx}
\usepackage{sectsty}
\usepackage{graphicx}
\usepackage{lastpage}
\usepackage[format=plain,justification=justified,singlelinecheck=false,font={stretch=1.125,small,sf},labelfont=bf,labelsep=space]{caption}
\usepackage{float}
\usepackage{fancyhdr}
\usepackage{fnpos}
\usepackage[english]{babel}
\addto{\captionsenglish}{%
  
}
\usepackage{array}
\usepackage{droidsans}
\usepackage{charter}
\usepackage[T1]{fontenc}
\usepackage[usenames,dvipsnames]{xcolor}
\usepackage{setspace}
\usepackage[compact]{titlesec}
\usepackage{hyperref}

\usepackage{epstopdf}
\usepackage{amsmath,amssymb}

\definecolor{cream}{RGB}{222,217,201}

\begin{document}

\pagestyle{fancy}
\thispagestyle{plain}
\fancypagestyle{plain}{
\renewcommand{\headrulewidth}{0pt}
}

\makeFNbottom
\makeatletter
\renewcommand\LARGE{\@setfontsize\LARGE{15pt}{17}}
\renewcommand\Large{\@setfontsize\Large{12pt}{14}}
\renewcommand\large{\@setfontsize\large{10pt}{12}}
\renewcommand\footnotesize{\@setfontsize\footnotesize{7pt}{10}}
\makeatother

\renewcommand{\thefootnote}{\fnsymbol{footnote}}
\renewcommand\footnoterule{\vspace*{1pt}%
\color{cream}\hrule width 3.5in height 0.4pt \color{black}\vspace*{5pt}} 
\setcounter{secnumdepth}{5}

\makeatletter 
\renewcommand\@biblabel[1]{#1}
\renewcommand\@makefntext[1]%
{\noindent\makebox[0pt][r]{\@thefnmark\,}#1}
\makeatother 
\renewcommand{\figurename}{\small{Fig.}~}
\sectionfont{\sffamily\Large}
\subsectionfont{\normalsize}
\subsubsectionfont{\bf}
\setstretch{1.125} 
\setlength{\skip\footins}{0.8cm}
\setlength{\footnotesep}{0.25cm}
\setlength{\jot}{10pt}
\titlespacing*{\section}{0pt}{4pt}{4pt}
\titlespacing*{\subsection}{0pt}{15pt}{1pt}

\fancyfoot{}
\fancyfoot[LE,RO]{\thepage}
\fancyhead{}
\renewcommand{\headrulewidth}{0pt} 
\renewcommand{\footrulewidth}{0pt}
\setlength{\arrayrulewidth}{1pt}
\setlength{\columnsep}{6.5mm}
\setlength\bibsep{1pt}

\makeatletter 
\newlength{\figrulesep} 
\setlength{\figrulesep}{0.5\textfloatsep} 

\newcommand{\topfigrule}{\vspace*{-1pt}%
\noindent{\color{cream}\rule[-\figrulesep]{\columnwidth}{1.5pt}} }

\newcommand{\botfigrule}{\vspace*{-2pt}%
\noindent{\color{cream}\rule[\figrulesep]{\columnwidth}{1.5pt}} }

\newcommand{\dblfigrule}{\vspace*{-1pt}%
\noindent{\color{cream}\rule[-\figrulesep]{\textwidth}{1.5pt}} }

\makeatother

\twocolumn[
  \begin{@twocolumnfalse}
\sffamily

\noindent\LARGE{\textbf{Experimental Study on Fracture Structure of Sumi-Wari}} \\

\noindent\large{Michiko Shimokawa,$^{\ast}$\textit{$^{a}$}, Lucas Goehring,\textit{$^{b}$}, Akie Kinoshita,\textit{$^{a}$}, Ludovic Pauchard,\textit{$^{c}$}
 and Hidetsugu Sakaguchi\textit{$^{d,e}$}} \\

\noindent\normalsize{Local variations in surface tension can induce complex fracture dynamics in thin interfacial films. Here, we investigate the fracture patterns that emerge when a localized surface-tension perturbation is applied to a sumi film supported on a water-glycerol subphase.  Sumi is a traditional Japanese carbon black ink, and this process, referred to as sumi-wari, produces aesthetically pleasing, star-shaped crack patterns with multiple spikes radiating from the perturbation site. The number of crack spikes increases with the viscosity of the subphase, controlled here by the addition of glycerol.  Atomic force microscopy measurements reveal that the effective stiffness of the sumi film decreases as glycerol concentration increases. This suggests a strong coupling between the subphase properties and the mechanics of the sumi film. To capture the dynamics of sumi-wari, a phenomenological model is outlined, based on an overdamped equation of motion for particles connected by breakable springs. Numerical simulations reproduce both the morphology and the experimental trends of sumi-wari: the number of cracks and their temporal evolution depend on the spring stiffness, mirroring the behavior observed for subphases with different viscosities. These findings demonstrate how the interplay between surface-tension gradients, subphase properties, and film mechanics governs local fracture and pattern formation in fluid-supported thin films.} \\


 \end{@twocolumnfalse} \vspace{0.6cm}

  ]

\renewcommand*\rmdefault{bch}\normalfont\upshape
\rmfamily
\section*{}
\vspace{-1cm}


\footnotetext{\textit{$^{a}$~Nara women's University, Kita-uoya Nishimachi, Nara, 630-8560, Japan. E-mail: shimokawa@cc.nara-wu.ac.jp}}
\footnotetext{\textit{$^{b}$~School of Science and Technology, Nottingham Trent University, Clifton Lane,
Nottingham, NG11 8NS, UK.}}
\footnotetext{\textit{$^{c}$~Fluides, Automatique et Syst\`emes Thermiques, Rue Andr\'e Rivi\`ere, 91405 Orsay cedex, France.}}
\footnotetext{\textit{$^{d}$~Kyushu University, 6-1 Kasuga-koen, Kasuga-shi, Fukuoka, 816-8580, Japan.}}
\footnotetext{\textit{$^{e}$~Nihon University, 3-11-3 Kanda-Surugadai, Chiyoda-ku, Tokyo, 101-8308, Japan.}}




\newcommand{\red}[1]{\textcolor{red}{#1}}

\newcommand{\bx}{{\bf x}}
\newcommand{\be}{{\bf e}}

\section{Introduction}
\begin{figure}[tbp]
\centering
  \includegraphics[width=\linewidth]{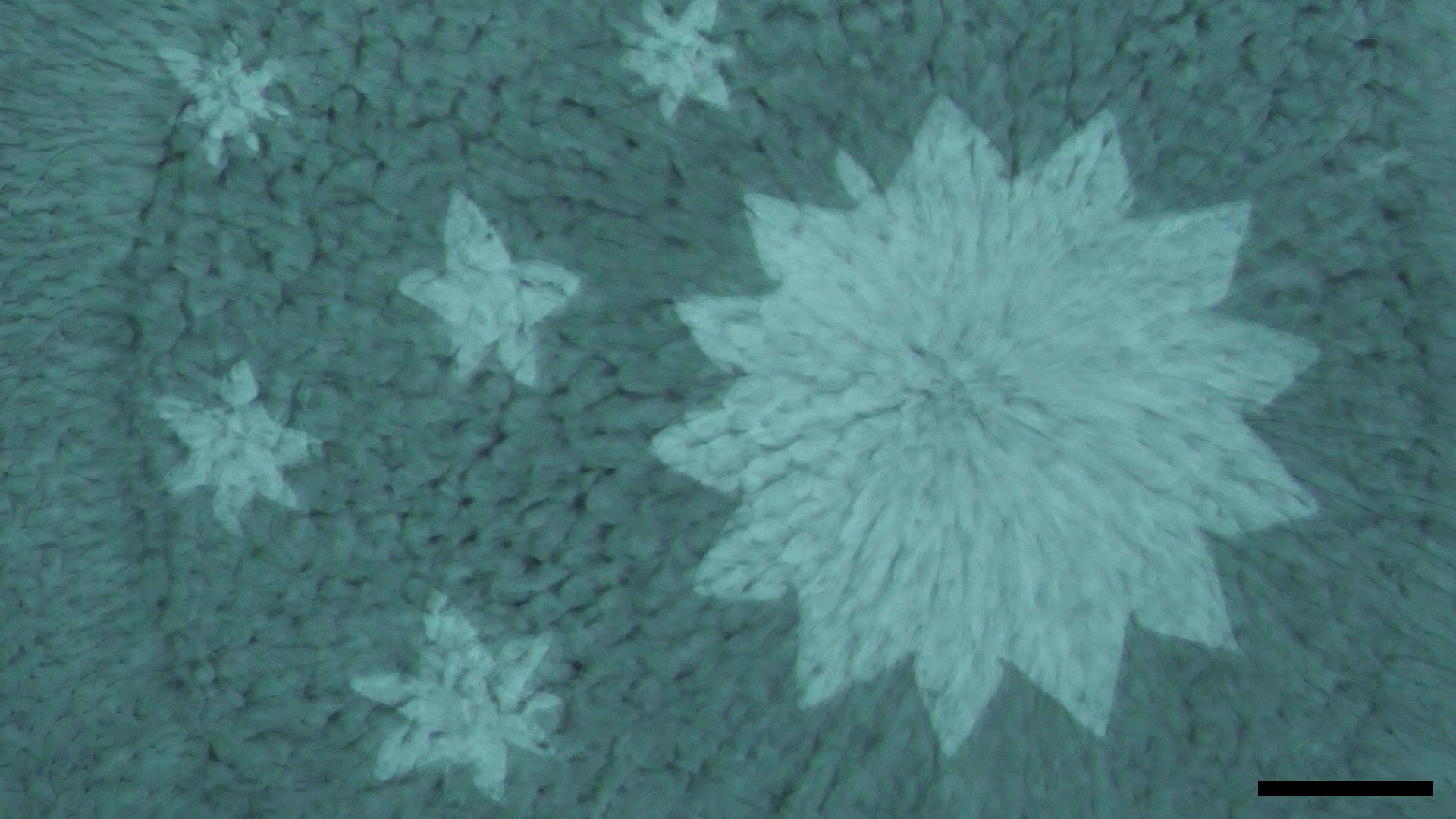}
  \caption{Sumi-wari.  Sumi ink contains hydrophobic carbon particles that can be easily spread into a thin, floating film.  Sumi-wari is made by breaking this film with a surfactant-laden needle, creating a star-shaped or sunburst crack pattern. Solid line in the image is a scale bar of 2.0 cm.}
  \label{fgr:sumi}
\end{figure}

Fracture and pattern formation in thin films are ubiquitous phenomena in soft and complex materials \cite{pattern1, balloon, Bacchin2018}. They arise from mechanical instabilities driven by gradients in stress, surface tension, or elastic modulus, and are observed in diverse systems ranging from drying colloidal suspensions and polymer coatings to biological membranes and Langmuir monolayers \cite{Ries1979,Lipp1998,colloid}. In such systems, crack patterns provide valuable insight into the underlying mechanical and interfacial properties, acting as visible signatures of stress relaxation and material heterogeneity \cite{pattern3}. The {\it sumi-wari} phenomenon -- the spontaneous formation of star-shaped cracks when a local perturbation is applied to a thin sumi film at the air-liquid interface -- offers a particularly simple and controllable model for studying such interfacial fracture dynamics. A typical pattern is shown in Fig. \ref{fgr:sumi}.  Although the term originates from traditional Japanese ink-painting techniques, sumi-wari can be viewed more generally as an interfacial instability akin to the cracking and folding seen in compressed Langmuir films \cite{langmuir1,langmuir2} or drying particle-laden layers \cite{Bacchin2018}. Studying sumi-wari thus provides a minimal experimental framework to probe how mechanical properties, such as film stiffness or subphase viscosity, govern the initiation and evolution of cracks in thin interfacial films.

The science of sumi-wari can be traced to a series of exploratory papers by Torahiko Terada in 1934 \cite{terada1,terada2,nakaya}. 
Sumi ink is made from hydrophobic soot (carbon black or lamp black) mixed with a collagen-based animal glue and water \cite{sumi,kimura}. 
A sumi film can be created by simply touching an ink-laden brush onto water. The ink will spread over the surface of the water in a matter of seconds, and form into a thin, floating film.  When a needle or toothpick coated with a detergent is inserted into a sumi film, cracks will develop.  These fractures are driven by the stresses induced in the film by differences in interfacial tension. The result is the spontaneous formation of a star-shaped pattern of cracks. Terada et al. \cite{terada1,terada2,nakaya} reported that copper ions in the subphase promote the aggregation of sumi colloids.  When the sumi ink spreads across a subphase containing copper ions, this aggregation leads to the spontaneous formation of a rigid sumi film. They further showed that the pattern of the sumi-wari depends on the elapsed time after the sumi ink spreads on the subphase: the number of the spikes on the pattern decreases with the age of the film. They argued that the solidification of the sumi film, due to the advanced colloidal aggregation, causes the pattern to change as a function of time. However, because the solidification occurred so rapidly, it was difficult to measure the physical properties of the sumi film quantitatively, even though its stiffness was a crucial factor in pattern formation. To address this limitation, we conducted experiments to control the colloid aggregation and alter the spatial average value of the stiffness of the sumi film by the addition of glycerol. The glycerol changes the viscosity of the solution, which can affect the dynamics of spreading and colloidal aggregation.

Similar experimental studies to sumi-wari, where two-dimensional fracture occurs due to the interfacial tension of a solution, have also been reported \cite{particle,carbon}. In these studies non-Brownian particles were packed onto the water surface to form a monolayer film, and surfactant was poured into a small space on the surface. The diameters of the particles in these experiments, on the order of $10^5$ nm, are much larger than the 10-150 nm of sumi particles \cite{sumi,kimura}.  In these experiments, crack propagation was observed to proceed by the creation of gaps between the particles. The pattern is the shape in which the cracks meander and is noticeably different from the pattern observed in our sumi-wari experiments. Thus, the pattern formation of sumi-wari cannot be explained in the framework reported in previous studies.

In this paper, we focus on the pattern formation of two-dimensional fracture phenomena in thin sumi films, with a thickness of approximately 100 nm, spread over the surface of a shallow solution. By building on Terada’s experiments \cite{terada1,terada2,nakaya}, the physical properties of the thin sumi film were varied, leading to the emergence of a diversity of sumi-wari patterns. We measured the macroscopic physical quantities as the basic physical properties of the sumi film, such as the height and the stiffness, using atomic force microscopy (AFM) measurements. Based on these experimental results, we proposed a phenomenological model for the surfactant-driven fracture of sumi films, i.e. sumi-wari. This paper presents detailed experimental results and discusses the dynamics of sumi-wari through a quantitative comparison between the experimental results and the mathematical model.

\section{Experimental method}

\begin{figure}[tbp]
\centering
  \includegraphics[width = \linewidth]{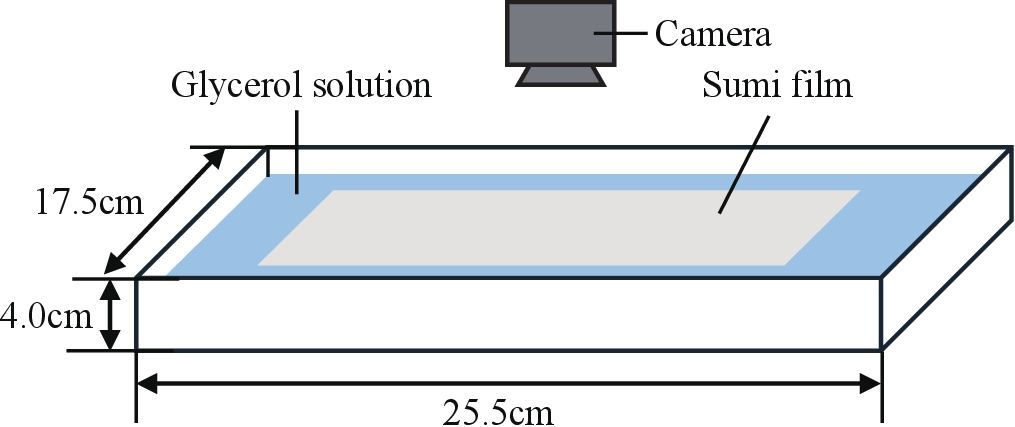}
  \caption{Schematic of the experimental setup.}
  \label{fgr:setup}
\end{figure}

The experimental method is sketched in Fig. \ref{fgr:setup}. We poured 150 ml of glycerol solution into a tray and then added 10 ml of alum solution. The alum solution was prepared by mixing 1 g of alum ($\rm{KAl(SO_4)_2\cdot12H_2O}$, Yaki Myoban, Tsukemoto) with 100 g of deionized water. 
In the place of copper salts used in the previous studies \cite{terada1,terada2,nakaya}, we used an alum solution to induce aggregation.  
The glycerol solution was prepared by mixing glycerol (Glycerol, Fuji film) with tap water. We used the glycerol concentration to control the viscosity of the solution \cite{glycerol}, which was measured by a tuning-fork vibration viscometer (SV-10A, A\&D Company). The solution in the tray had a final alum concentration of $6.2 \times 10^{-4}$ g/ml and a thickness of approximately 3 mm. This is shallower than the water layers used in previous experiments on surfactant-driven fracture of particle rafts \cite{particle,carbon}; we found that using a thin layer reduced the influence of any fluid motion occurring under the film during the sumi-wari dynamics. After mixing the glycerol and alum solutions together, a brush coated with sumi ink (Bokuju BO1020, KAIMEI) was gently dipped into the surface of the solution. The sumi, also known as Chinese ink, is composed of soot, glue, and water \cite{sumi}. As soon as the brush touched the surface of the solution, the ink spread rapidly across the surface. The system was then left for $15~{\rm minutes}$, during which time a rigid sumi film formed by the aggregation of the soot particles, i.e. the colloidal carbon. The sumi film floated stably on the solution throughout the experiments. Surfactant solution was also prepared by mixing soap (Daidokoroyou Senzai, KAO) with deionized water. Its surface tension was measured as $26 \pm 0.6$ mN/m using a surface tension meter (Kyowa Kaimen Kagaku, DMs401). Finally, the tip of a toothpick was covered with the surfactant solution and inserted into the sumi film. This caused the spontaneous formation of sumi-wari cracks. The dynamics of the sumi-wari were recorded with a camera (HC-X1500, Panasonic) positioned above the container.

\section{Experimental results}

\subsection{Viscosity dependence of sumi-wari patterns}


\begin{figure*}[tbp]
\centering
  \includegraphics[width=\linewidth]{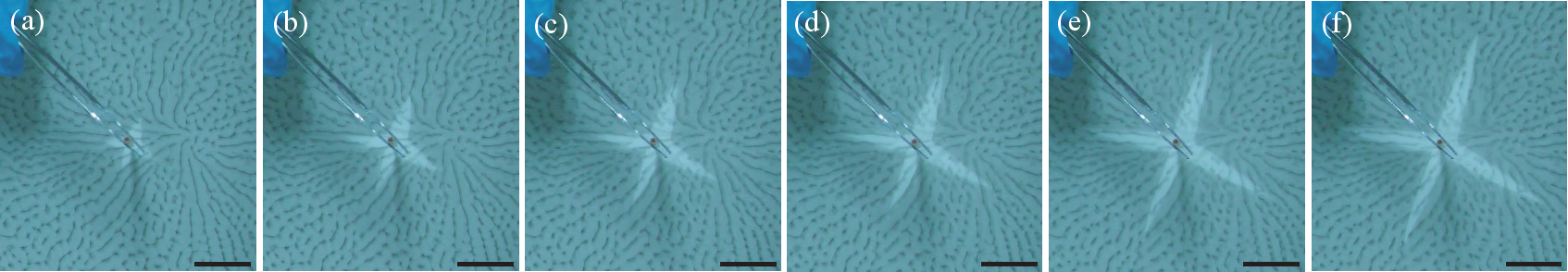}
  \caption{Sumi-wari patterns observed in experiments with subphase viscositiy $\eta =1.0 ~{\rm mPa}\cdot{\rm s}$ ($c = 0 ~\%$) at (a) $t=0.1~{\rm s}$, (b) $0.3~{\rm s}$, (c) $0.6~{\rm s}$, (d) $1.0~{\rm s}$, (e) $1.5~{\rm s}$ and (f) $2.0~{\rm s}$. 
The solid lines below these images are scale bars of $2.0~{\rm cm}$.}
  \label{fgr:time}
\end{figure*}

Figure~\ref{fgr:time} shows the time development of a sumi-wari pattern observed in an experiment using a solution with a viscosity $\eta =1.0 ~{\rm mPa}\cdot{\rm s}$ ($c = 0 ~\%$), where $c$ is the weight percentage of glycerol in the solution.
As soon as we insert the surfactant-containing toothpick into the sumi film at (a) $t=0$, the sumi-wari pattern, which resembles a sunburst or star-shaped opening, appears. The area of the sumi-wari pattern increases with time, as shown in panels (b)-(d), and appears to be driven by the difference in interfacial tension between the inside of the open area, including the toothpick coated with detergent, and the outside of the still-intact sumi film.   The growth of the sumi-wari pattern stops after some time, however, with a  duration that depends on the viscosity of the subphase solution ((e), (f)). We also observed the growth dynamics using a high-speed camera (HAS220, DITECT). The spikes of the sumi-wari pattern appeared earlier than 1/1000 s after the toothpick was inserted. The number of spikes remained constant during the pattern growth.
Similar tendencies were observed in experiments with different viscosities.

\begin{figure}[tbp]
\centering
  \includegraphics[width=\linewidth]{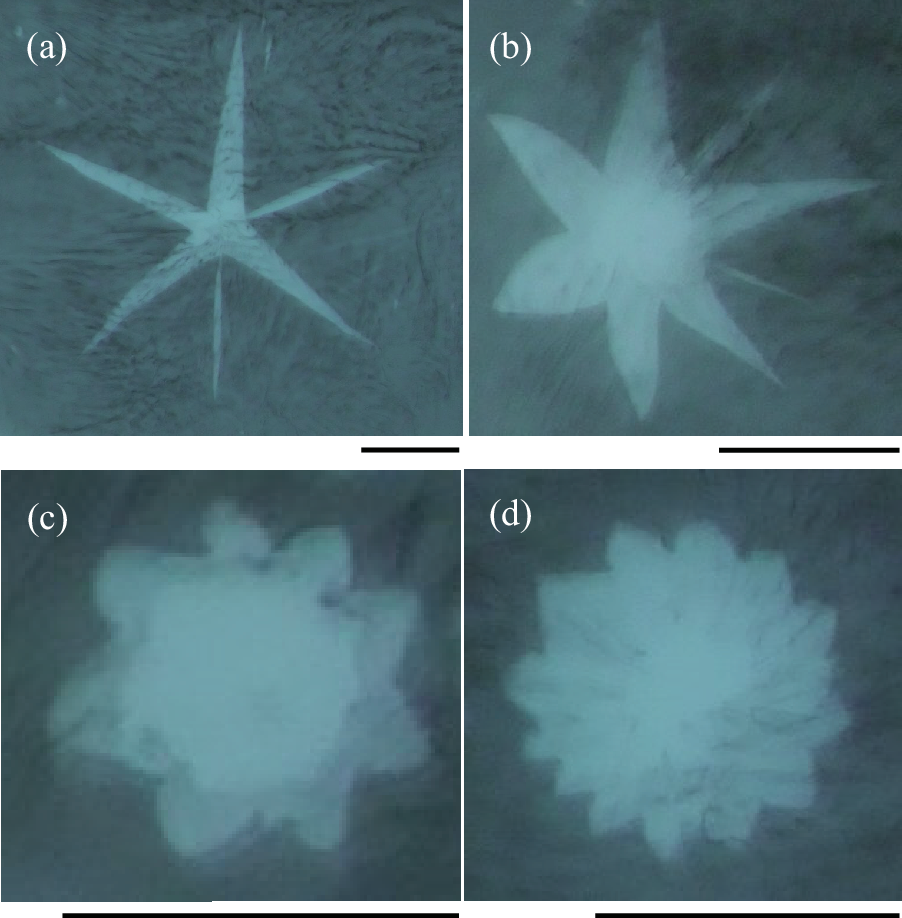}
  \caption{Sumi-wari patterns observed in experiments of several different subphase viscosities, with (a) $\eta =1.0 ~{\rm mPa}\cdot{\rm s}$ ($c = 0 ~\%$) at $t=2.7~{\rm s}$, (b) $1.5~{\rm mPa}\cdot{\rm s}$ ($c = 15~\%$) at $t=2.0~{\rm s}$, (c) $2.4~{\rm mPa}\cdot{\rm s}$ ($c = 30~\%$) at $t=2.5~{\rm s}$ and (d) $3.1~{\rm mPa}\cdot{\rm s}$ ($c = 35~\%$) at $t=2.8~{\rm s}$.  Here, $t=0$ is a time when the surfactant-laden toothpick was inserted. The solid lines below these images are scale bars of $3.0~{\rm cm}$.}
  \label{fgr:pattern}
\end{figure}

\begin{figure}[tbp]
\centering
  \includegraphics[width=\linewidth]{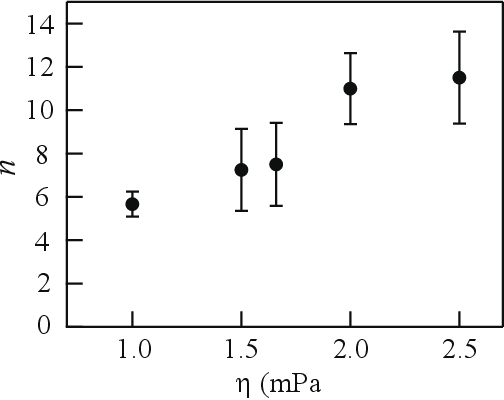}
  \caption{Relationship between the viscosity $\eta$ of the solution under the film and the number $n$ of spikes of the sumi-wari patterns. Each value of $n$ is the average from at least seven independent experimental trials.
The error bars show the standard deviation.}
  \label{fgr:n}
\end{figure}

The sumi-wari pattern consists of several spikes or cracks. 
We varied the amount of glycerol $c$ in the solution under the sumi film and investigated the viscosity dependence of the sumi-wari pattern. Figure \ref{fgr:pattern} shows the patterns observed at different viscosities of (a) $\eta =1.0 ~{\rm mPa}\cdot{\rm s}$ ($c = 0~\%$), (b) $1.5~{\rm mPa}\cdot{\rm s}$ ($c = 15~\%$), (c) $2.4~{\rm mPa}\cdot{\rm s}$ ($c = 30~\%$) and (d) $3.1~{\rm mPa}\cdot{\rm s}$ ($c = 35~\%$). The shape became more rounded, and the number of spikes $n$ increased, as the viscosity $\eta$ increased. Figure \ref{fgr:n} shows how the average number of spikes depends on the subphase viscosity, over the range of $1.0~{\rm mPa}\cdot{\rm s}$ $\leq\eta\leq2.5 ~{\rm mPa}\cdot{\rm s}$. The average values of $n$, shown in Fig. \ref{fgr:n}, were obtained from at least seven different experiments for each viscosity condition.

\subsection{Dynamics and growth of sumi-wari patterns}

\begin{figure}[tbp]
\centering
  \includegraphics[width=\linewidth]{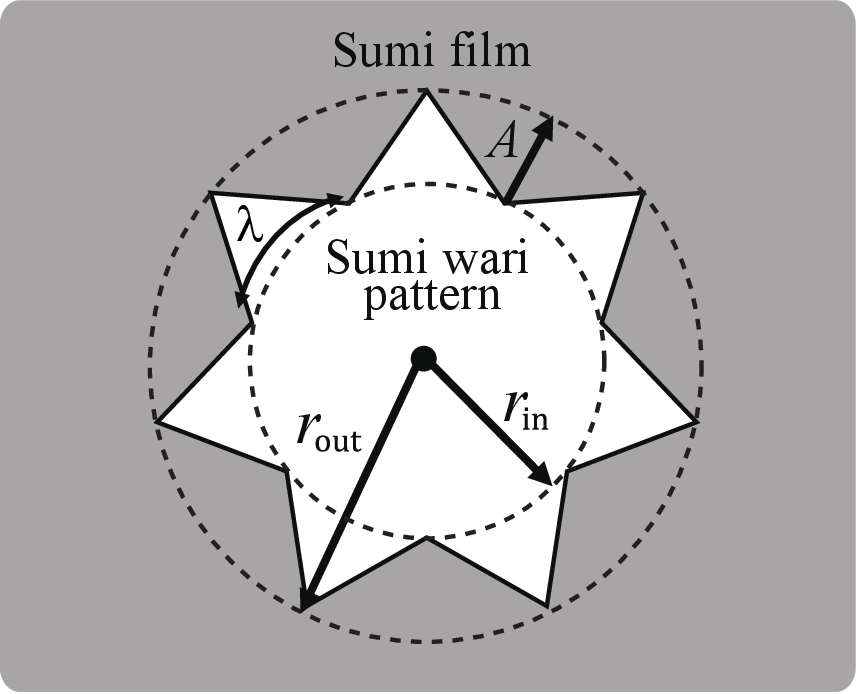}
  \caption{Schematic of the radii of the inner circle $r_{\rm in}$ and the outer circle $r_{\rm out}$ of the sumi-wari pattern. The amplitude and wavelength of the spikes are represented by $A$ and $\lambda$.}
  \label{fgr:rin2}
\end{figure}

The growth of the cracks of a sumi-wari pattern can be measured through changes in the sizes of the circles that define an inner and outer radius of the cracked area, $r_{\rm in}$ and $r_{\rm out}$, as in Fig. \ref{fgr:rin2}.  Using images captured by the camera (HC-X1500, Panasonic) with a time resolution of 1/60 s (e.g., Fig.~\ref{fgr:time}), we investigated how these radii developed over time, to quantify the dynamics of sumi-wari pattern formation.  From these images we also extracted the characteristic amplitude of the spikes, $A=r_{\rm out} - r_{\rm in}$, and their wavelength, $\lambda = 2 \pi r_{\rm in}/n$, where $n$ denotes the number of spikes of the pattern.

\begin{figure}[tbp]
\centering
  \includegraphics[width=\linewidth]{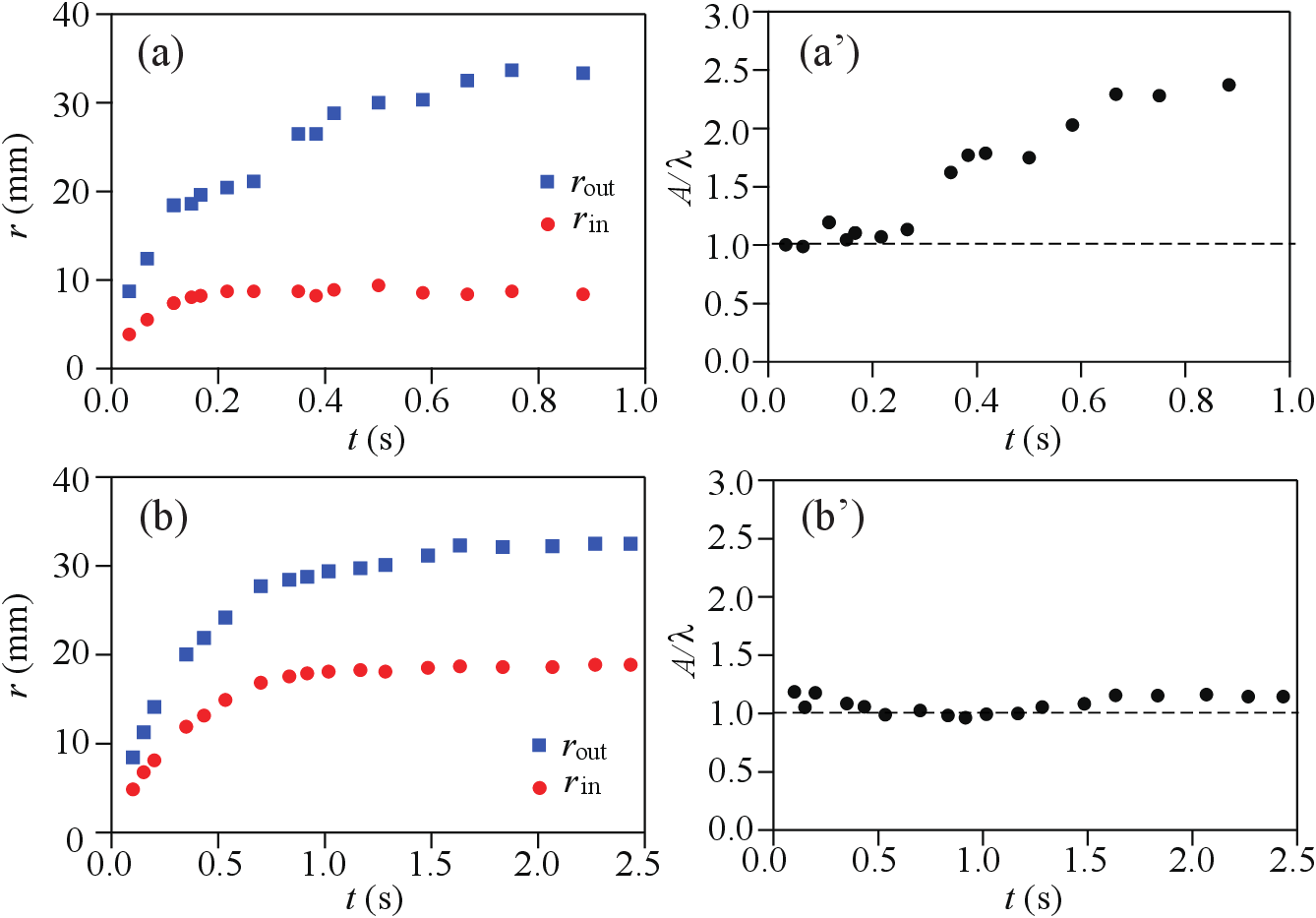}
  \caption{(a), (b) Time series of the radii of $r_{\rm in}$ and $r_{\rm out}$ of the sumi-wari pattern in experiments with (a) $1.0 ~{\rm mPa}\cdot{\rm s}$ ($c = 0~\%$) and (b) $2.5 ~{\rm mPa}\cdot{\rm s}$ ($c = 30~\%$), respectively. Time is represented by $t$, where $t=0$ is the time when the toothpick was inserted into the film. Closed squares and circles represent the data of $r_{\rm out}$ and $r_{\rm in}$ , respectively. (a'), (b') Relationship between $t$ and $A/\lambda$, where $\lambda$ represents the wavelength of the spikes and $A=r_{\rm out} - r_{\rm in}$. Dashed lines represent $A/\lambda\sim 1$. The data of (a') and (b') correspond to (a) and (b), respectively.}
  \label{fgr:r-1}
\end{figure}

Figure \ref{fgr:r-1} shows how $r_{\rm in}$ and $r_{\rm out}$ develop over time $t$ in experiments of (a) $1.0~{\rm mPa}\cdot{\rm s}$ ($c = 0~\%$) and (b) $2.5 ~{\rm mPa}\cdot{\rm s}$ ($c = 30~\%$), where $t=0$ is the time at which a toothpick was inserted into the sumi film. The closed circles and squares are data for $r_{\rm in}$ and $r_{\rm out}$, respectively. For the lower-viscosity case, Fig. \ref{fgr:r-1}(a) shows that $r_{\rm in}$ and $r_{\rm out}$ increase together up to a time $t\simeq 0.3$ s.  After this time the crack tips continue to grow, and hence $r_{\rm out}$ continues to increase, but the inner radius $r_{\rm in}$ saturates at a fixed size. In contrast, for the higher-viscosity case shown in Fig. \ref{fgr:r-1}(b), $r_{\rm in}$ and $r_{\rm out}$ increase with time and saturate together at a similar time. 

The differences between the growth of the inner and outer limits of the spikes become more apparent in Figs. \ref{fgr:r-1} (a') and (b'). These show the relationship between $t$ and $A/\lambda$ for (a') $1.0 ~{\rm mPa}\cdot{\rm s}$ ($c = 0~\%$) and (b') $2.5 ~{\rm mPa}\cdot{\rm s}$ ($c = 30~\%$), and the data correspond to those of panels (a) and (b). As shown in Fig. \ref{fgr:r-1} (a'), $A/\lambda\sim 1$ remains constant when $t \lesssim 0.3$, and $A/\lambda > 1$ for $t \gtrsim 0.3$. Since the maintenance of $A/\lambda$ is consistent with the growth of a self-similar shape, the data suggests that this sumi-wari pattern initially grew to maintain the same shape, with the spikes only developing independently after $t \simeq 0.3$. 
This behavior is also confirmed in Fig.~\ref{fgr:time}.
The data obtained from the high viscosity experiments (Fig. \ref{fgr:r-1} (b')) shows $A/\lambda\sim 1$ at all times. Thus, it shows that the sumi-wari pattern grew to maintain a similar shape, at all stages of its development. The maintenance of similar shape on the sumi-wari pattern depends on the viscosity of the experimental condition.

\subsection{AFM measurements of the properties of sumi films}

Atomic force microscopy (AFM, Dimension Icon, Bruker) was used to investigate the physical properties of sumi films.  The resulting data were analysed in Gwyddion \cite{Gwyddion}. Four types of films were used for these measurements, in order to examine the effects of the underlying fluid's viscosity (i.e. the amount of glycerol) on the film properties, including its stiffness, thickness and surface roughness.  The films of the patterns shown in Fig. \ref{fgr:pattern} are very soft and were found to be difficult to transfer onto a rigid substrate for AFM measurement. As such, for the AFM tests the alum concentration and the time allowed for film formation were increased. The amount of alum in the solution was $1.2\times 10^{-3}$ g/ml (rather than $6.2\times 10^{-4}$ g/ml) and, after spreading, the sumi film was left undisturbed for over $15~{\rm hours}$ (rather than $15~{\rm minutes}$).

\begin{figure}[tbp]
\centering
  \includegraphics[width=\linewidth]{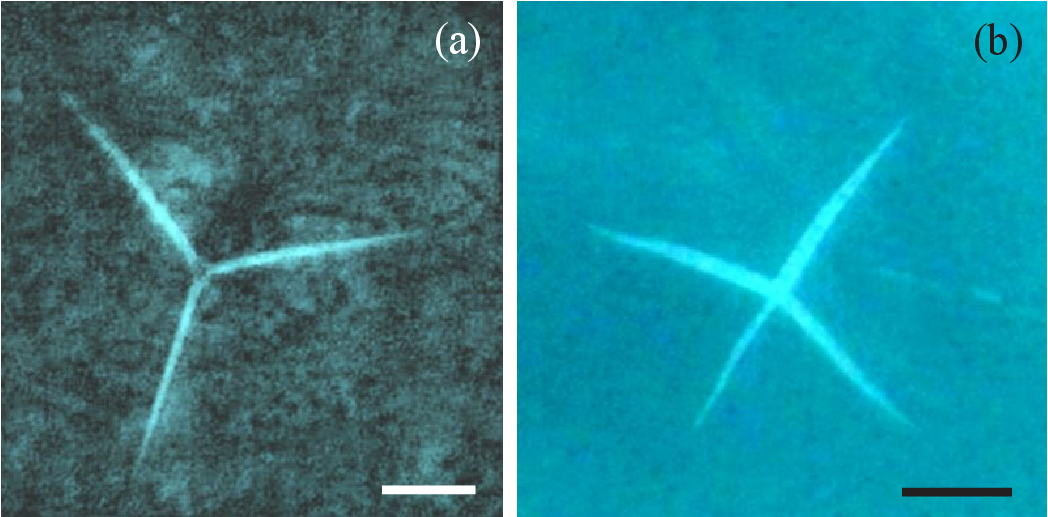}
  \caption{Sumi-wari patterns formed on the surfaces of (a) water with a viscosity of $1.0~{\rm mPa}\cdot{\rm s}$ ($c = 0~\%$) and (b) a glycerol solution with a viscosity of $1.5~{\rm mPa}\cdot{\rm s}$ ($c = 15~\%$). The sumi films were left on the surface of the solutions for $15~{\rm h}$ before inserting a toothpick covered with surfactant. The solid lines in the figures represent $20~{\rm mm}$ scale bars.}
  \label{fgr:15h}
\end{figure}

We confirmed that sumi-wari patterns formed under the conditions used for the AFM experiments.  This is important, as these patterns are known to be influenced by both the aggregation time, and the concentration of salts in the subphase \cite{terada1, nakaya}. In figure \ref{fgr:15h}, panels (a) and (b) show the patterns now observed over a water bath with $\eta=1.0 ~{\rm mPa}\cdot{\rm s}$ and a water-glycerol bath with $\eta=1.5 ~{\rm mPa}\cdot{\rm s}$, respectively.   The number of spikes in these patterns is smaller than those explored in Sections 3.1 and 3.2.  There are fewer spikes, and the spikes are narrower than those shown in Fig. \ref{fgr:pattern}, but there is a similar tendency for the number of spikes to increase with the viscosity of the subphase: for example, $n = 3.3$ (standard deviation $\sigma=0.4$) for $\eta=1.0 ~{\rm mPa}\cdot{\rm s}$ and $n = 3.8$ ($\sigma=0.8$) for $\eta=1.5 ~{\rm mPa}\cdot{\rm s}$.  

Sumi films were transferred onto glass microscope slides by gently touching a clean slide to the floating film. The films were then kept covered with water, to avoid any structural changes that could be caused by desiccation. We performed quantitative nanomechanical mapping (QNM \cite{r1,r2}) on the films using a liquid immersion environment (fluid cell) for the AFM probe. We used RTESPA-150 tapping-mode probes with a nominal spring constant of 5 N/m, which are optimal for measuring surfaces with a Young's modulus of between approximately 5--500 MPa. The sensitivity (typical value, 45 nm/V) and the spring constant (typical value, 3.7 N/m) of each cantilever were calibrated against a sapphire plate. We employed a relative indentation method, where we first indented approximately 5 nm into a PDMS test surface with a known Young's modulus of $E=3.5$ MPa (reduced modulus of $E^*=E/(1-\nu^2)=4.67$ MPa, given a Poisson ratio of $\nu=0.5$ for PDMS \cite{pdms}). Using this method, the effective tip radius of the probe was adjusted until the measured modulus matched its expected value. We then proceeded to probe the sumi films, adjusting the peak force setpoint (typical value, 3 nm) to ensure a similar indentation depth of between 2--10 nm.

\begin{figure*}[tbp]
\centering
  \includegraphics[width=\linewidth]{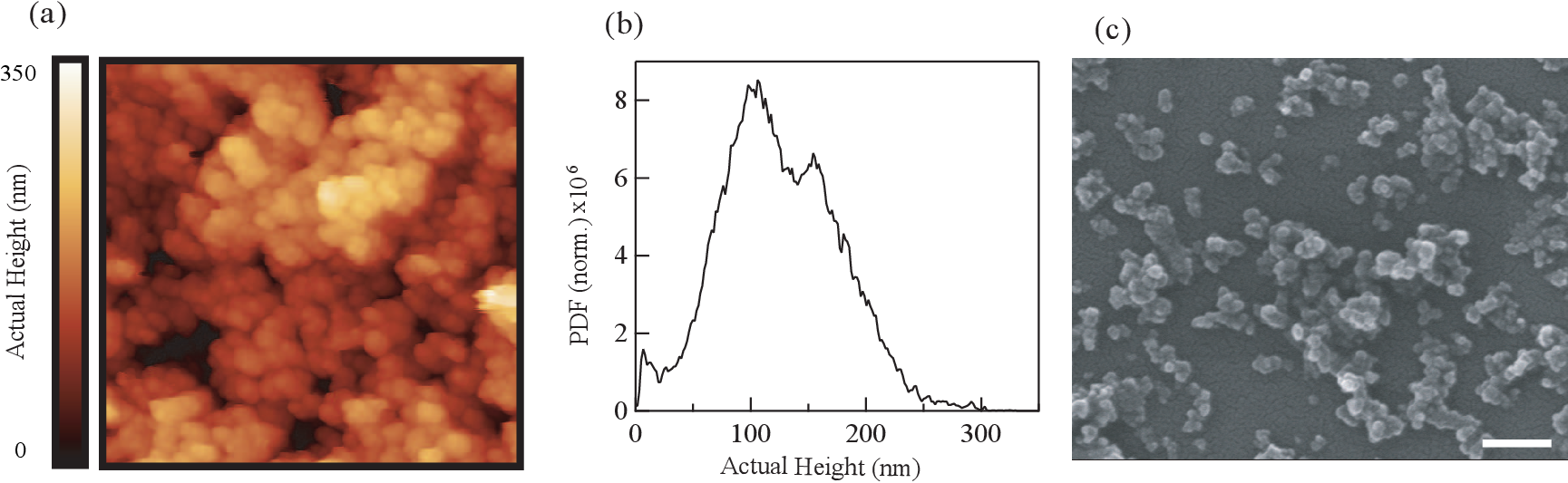}
  \caption{Structure of sumi films. (a) Actual height profiles of a film prepared over water with a viscosity of $1.0~{\rm mPa}\cdot{\rm s}$, as measured by atomic force microscopy. The measurement area is $1.5 ~\mu {\rm m}\times 1.5 ~\mu{\rm m}$, and the color bar or $z$ axis represents the height of the sumi film at any position, where $z=0$ is referenced to the height of the flat glass substrate on which the film is supported (fixed by three-point plane fitting at positions of gaps in the film). (b) Distribution of the heights over the area shown in (a), and representing the film thickness. (c) Picture of the dried sumi films captured using scanning electron microscopy. The solid line gives a $300~{\rm nm}$ scale bar.}
 \label{fgr:actual_height1}
\end{figure*}

The AFM measurements also map out the height profiles of the sumi films, allowing access to details of their surface roughness. Figure \ref{fgr:actual_height1}(a) shows the surface condition of a sumi film formed on the surface of tap water with alum, over an area of 1.5 $\mu$m$\times$ 1.5 $\mu$m. Although generally appearing as a dense film composed of colloidal aggregates, there are occasionally small gaps or holes through which the supporting glass slide can be seen. These were used to level the image by a three-point plane fit, so that the $z$ axis (color bar) in Fig. \ref{fgr:actual_height1}(a) represents the absolute value of the height of the film's surface above the slide. The height distribution for this film is given in Fig. \ref{fgr:actual_height1}(b).  It is a symmetrical distribution around a peak value of approximately 100 nm (mean 113 nm, standard deviation 54 nm), which we take to be characteristic of the sumi film thickness. Individual particles of colloidal carbon can also be seen in the AFM images, with a typical size of 50--100 nm. This is consistent with prior characterization of sumi inks, which suggests that the diameter of the primary particles should be in the range of 10--150 nm\cite{terada2,sumi}. To investigate this further, we performed scanning electron microscopy (SEM, JSM-7100F LV FEG) on dried sumi films, which were sufficiently broken up to allow individual colloidal particles to be seen, as in Fig. \ref{fgr:actual_height1}(c). The sizes of these grains were measured by hand in ImageJ \cite{R4}, and showed a mean particle diameter of 46 nm with a standard deviation of 9 nm. This suggests that the sumi films are only 2--3 particles thick. Given that the sumi-wari pattern extends horizontally over several centimeters, these results support an interpretation of the pattern formation process of sumi-wari as, effectively, two-dimensional fracture dynamics.

\begin{figure}[tbp]
\centering
  \includegraphics[width=\linewidth]{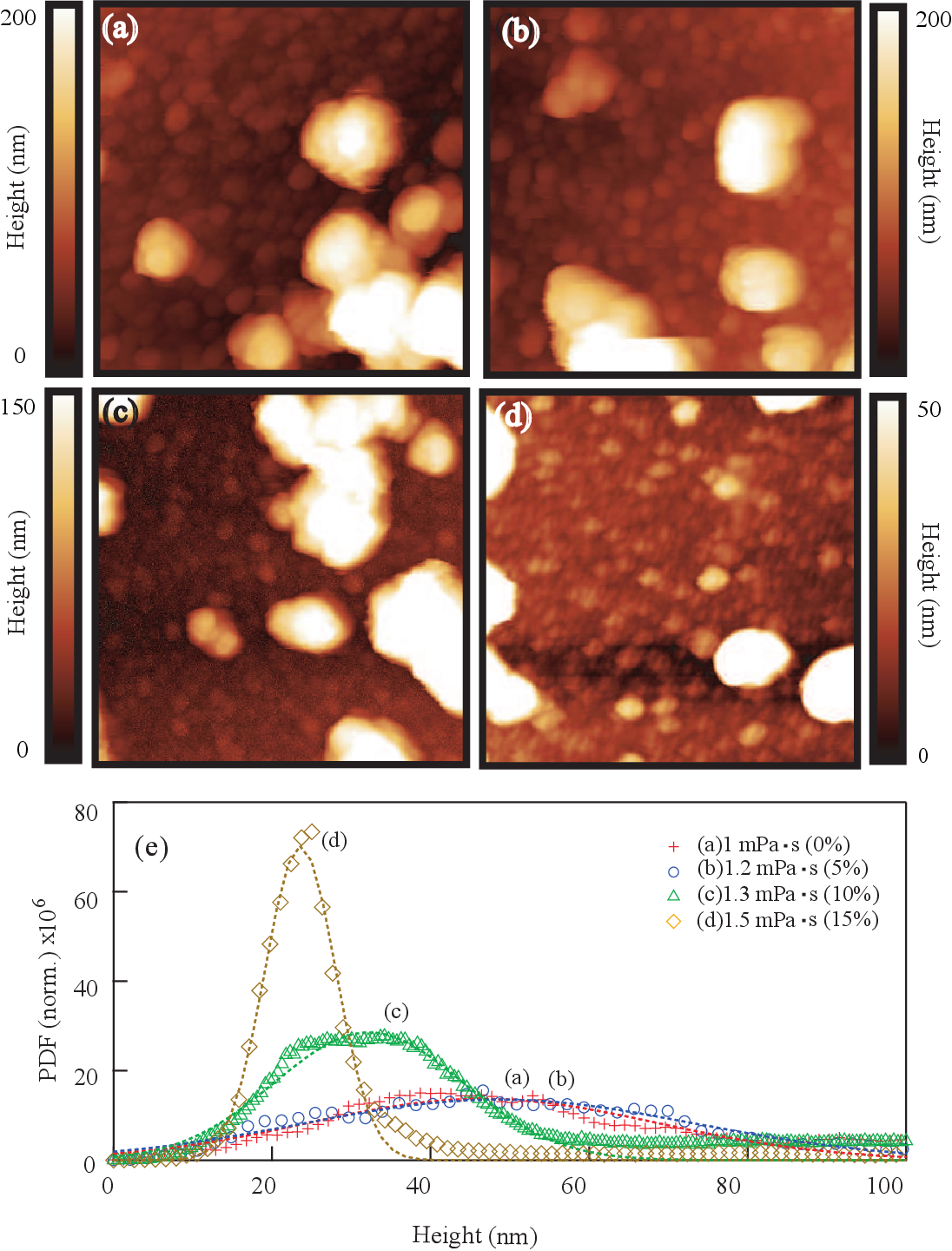}
  \caption{Structure of sumi films. (a-d) Height profiles of films prepared over (a) water with a viscosity of $1.0~{\rm mPa}\cdot{\rm s}$ ($c = 0~\%$), and (b-d) water-glycerol mixtures with a viscosity of (b) $1.2~{\rm mPa}\cdot{\rm s}$ ($c = 5~\%$), (c) $1.3~{\rm mPa}\cdot{\rm s}$ ($c = 10~\%$) and (d) $1.5~{\rm mPa}\cdot{\rm s}$ ($c = 15~\%$), as measured by atomic force microscopy. The measurement area is $1.0 ~\mu{\rm m}\times 1.0 ~\mu{\rm m}$ and the color bar or $z$ axis represents the height of the sumi sheet at the position, where $z=0$ is set to the lowest point measured on the surface of the film.  (e) Distribution of the heights of the areas shown in (a-d). Dashed lines represent fits to a Gaussian distribution.}
  \label{fgr:height-2}
\end{figure}

Figure \ref{fgr:height-2} shows the viscosity-dependence of the surface condition of sumi films. These films were formed over four solutions with (a) $\eta=1.0 ~{\rm mPa}\cdot{\rm s}$ ($c=0~\%$), (b) $1.2~{\rm mPa}\cdot{\rm s}$ ($c=5~\%$), (c) $1.3~{\rm mPa}\cdot{\rm s}$ ($c=10~\%$) and (d) $1.5~{\rm mPa}\cdot{\rm s}$ ($c=15~\%$), respectively. The measurement area is $1.0 ~\mu {\rm m}\times 1.0 ~\mu {\rm m}$, in each case. As these films were more continuous we could not set an absolute reference plane for the underlying glass plate, and instead the zero-point of the height is taken as the lowest point measured on the film. The color axes of the images have also been restricted, to highlight the differences in the film structure caused by changes in $\eta$. A characteristic of the sumi films, independent of $\eta$, is the presence of a few larger aggregates embedded in an otherwise very flat film. Close inspection of the backgrounds in Fig. \ref{fgr:height-2} shows that smaller particles have aggregated to form the continuous surface of these films. These particles have a similar size to those of the primary or individual sumi particles shown in Fig. \ref{fgr:actual_height1} (c). It is therefore considered that the flat film, below the large aggregates, is constructed by a dense packing of individual sumi colloids, largely without holes. Interestingly, the addition of glycerol to the subphase prompts the film to become flatter and much more uniform in height. This tendency is demonstrated by the effect of viscosity on the height distributions shown in Fig. \ref{fgr:height-2} (e). These differences are highlighted by the standard deviation $\sigma$ when we fit the data around the peak value with a Gaussian distribution. These fits are represented by dashed lines in Fig. \ref{fgr:height-2}(e). The roughness $\sigma$ of the films are 22.0 nm ($\pm 0.7$) for the case shown in Fig. \ref{fgr:height-2}(a), 24.7 nm ($\pm 0.6$) for Fig. \ref{fgr:height-2}(b), 12.0 nm ($\pm 0.3$) for Fig. \ref{fgr:height-2}(c) and only 4.5 nm ($\pm 0.1$) for Fig. \ref{fgr:height-2}(d).

\begin{figure}[tbp]
\centering
  \includegraphics[width=\linewidth]{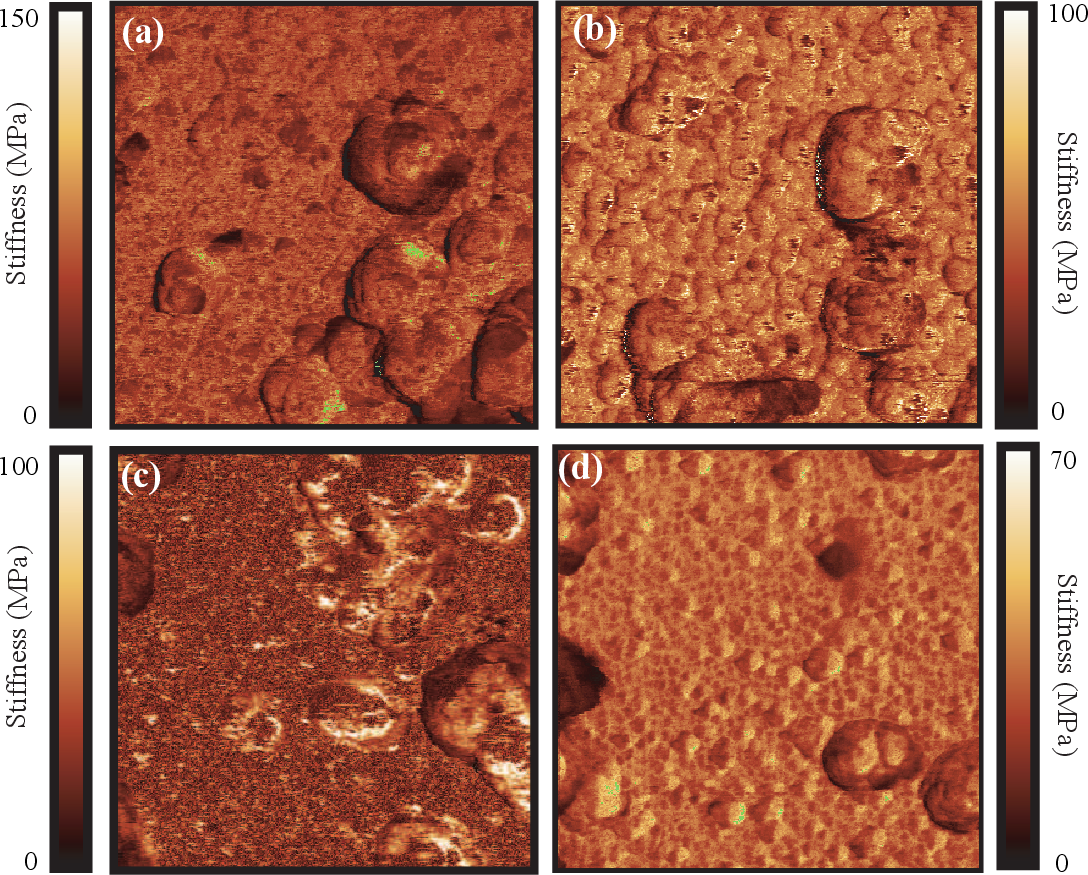}
  \caption{Stiffness of sumi films, as measured by atomic force microscopy. In each case the measurement area is $1.0 ~\mu{\rm m}\times 1.0 ~\mu{\rm m}$, and the color bar on $z$ axis represents the stiffness at each position. The sumi films were prepared over (a) a mixture of alum and water with a viscosity of $1.0~{\rm mPa}\cdot{\rm s}$ ($c = 0~\%$), (b-d) alum-water-glycerol mixtures with viscosities of (b) $1.2~{\rm mPa}\cdot{\rm s}$ ($c = 5~\%$), (c) $1.3~{\rm mPa}\cdot{\rm s}$ ($c = 10~\%$) and (d) $1.5~{\rm mPa}\cdot{\rm s}$ ($c = 15~\%$).}
  \label{fgr:stiffness-3}
\end{figure}

\begin{figure}[tbp]
\centering
  \includegraphics[width=\linewidth]{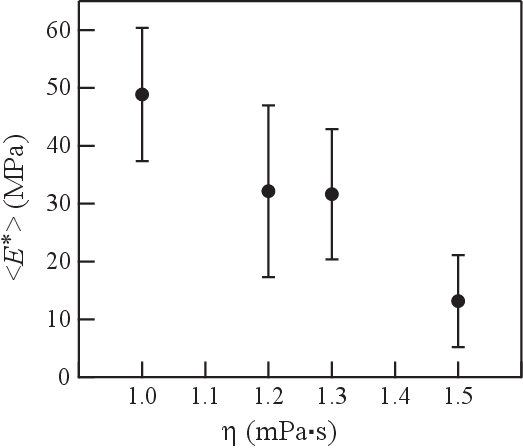}
  \caption{Average values $\langle E^\ast \rangle$ of the sumi film stiffness, depending on the viscosity $\eta$ of the subphase solution.  Error bars show the standard deviations of measurements for each experimental condition.}
  \label{fgr:actual_stiffness-2}
\end{figure}

We investigated the stiffness of the sumi films formed under the above conditions using Brucker's PeakForce QNM software, which fits the Derjaguin, Muller, Toprov model of deformation \cite{R3} to the retraction branch of the force-distance curves obtained by the indentation at each point of the AFM image. By this method the reduced modulus, $E^*=E/(1-v^2)$, is obtained by tapping the probe into the sumi film. This modulus is proportional to its Young's modulus, $E$, if the Poisson ratio $\nu$ in the film is constant. Figure \ref{fgr:stiffness-3} shows the results obtained for the sumi films formed over solutions with (a) $\eta=1.0 ~{\rm mPa}\cdot{\rm s}$ ($c=0~\%$), (b) $1.2~{\rm mPa}\cdot{\rm s}$ ($c=5~\%$), (c) $1.3~{\rm mPa}\cdot{\rm s}$ ($c=10~\%$) and (d) $1.5~{\rm mPa}\cdot{\rm s}$ ($c=15~\%$). The measurement area is $1.0 ~\mu {\rm m}\times 1.0 ~\mu{\rm m}$, in each case, and the color axes represent the values of $E^*$. The measured samples for panels (a)-(d) correspond to those in Figs. \ref{fgr:height-2}(a)-(d), respectively. The images of $E^*$ have similar spatial structures to the height images shown in Figs. \ref{fgr:height-2}(a)-(d), which gives confidence that the measurements of $E^*$ are accurately probing variations in local stiffness and, for example, highlighting differences between particles. As shown in Fig. \ref{fgr:stiffness-3}, the spatial pattern of $E^*$ of the sumi film prepared over a water-alum solution is more heterogeneous than those over the glycerol-alum solutions. Figure \ref{fgr:actual_stiffness-2} shows the relationship between the viscosity $\eta$ and $\langle E^*\rangle$, where $\langle E^\ast \rangle$ is the average value of $E^*$ obtained from at least seven different AFM images for each viscosity condition, and where the error bars show the standard deviation in the stiffness as measured between different images. The data show that the addition of glycerol to the solution lowers the stiffness of the sumi films.

Generally, the AFM images of the sumi films in Figs. \ref{fgr:height-2} and \ref{fgr:stiffness-3} demonstrate that the addition of glycerol to the subphase leads to systematic variations in the properties of the films: they become flatter and softer. A more detailed understanding of how the structure and the stiffness of these films change under different conditions, such as with alum concentration, is left for future study.

\section{Phenomenological model}

In this section, we consider the dynamics of the pattern formation of sumi-wari from the perspective of a phenomenological model, similar to that used to study desiccation-driven fracture in thin layers \cite{Kitsune1999}. The experimental results in the previous section show that the sumi-wari pattern and the stiffness of the sumi film depend on the amount of glycerol. Based on our experimental results, we propose a phenomenological model under the assumption that a sumi film is well-approximated by a thin layer of a brittle, elastic material.

The phenomenological model is governed by an overdamped equation of the motion of particles coupled together with breakable springs. The initial positions of the particles were determined by a Voronoi tessellation and its relaxation. More specifically, $N=10201$ particles were randomly scattered in a square of size $[-50,50]\times [-50,50]$. The average distance between neighboring particles is thus approximately 1, which corresponds to the average distance between the neighboring colloidal particles in the sumi film. Voronoi polygons were then constructed using the perpendicular bisectors of the line segments connecting the random points. From this, the Delaunay triangulation can be constructed by connecting all pairs of points in neighboring Voronoi cells. The result is a random network, but the variance of the distance between the neighboring particles is rather large. To reduce the variance, the position of each point in the Voronoi diagram was moved to the centre of  mass of its corresponding polygon. After moving all the particles, a new Voronoi tessellation was constructed. After 20 repetitions of this relaxation process, the particle positions were fixed. We used these random positions as the initial positions of the particles in the dynamical simulation, with the corresponding Delaunay triangulation defining the initial network of springs linking the particles together.

\begin{figure}[tbp]
\centering
  \includegraphics[width=\linewidth]{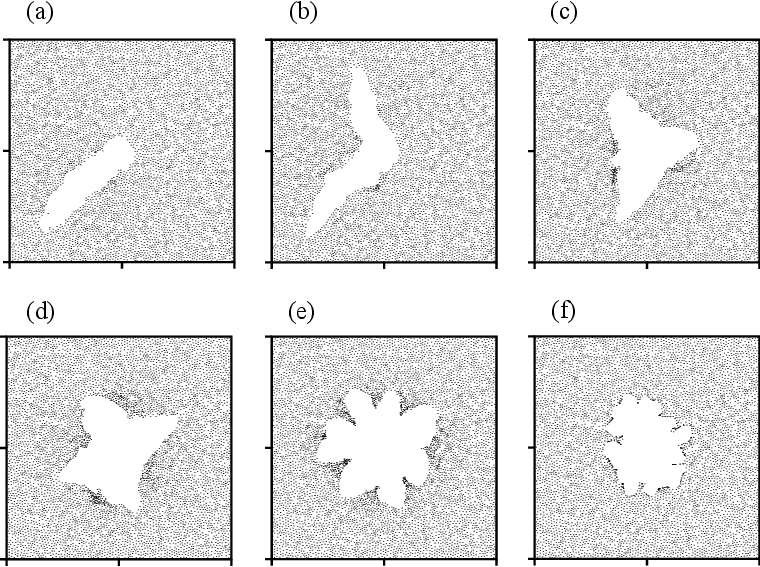}
  \caption{Sumi-wari pattern for values of (a) $k=4$, (b) $3$, (c) $1$, (d) $0.5$, (e) $0.2$, (f) $0.03$, and where $F=4$ and $s=0.2$. Snapshots at different times are chosen. Particle positions located in the region of $[-40, 40]\times[-40, 40]$ are plotted with dots.}
  \label{fgr:fig80}
\end{figure}

The initial position for the $i^{\rm th}$ particle is denoted as $x_{i0}$.  It is connected to a neighboring point $j$ of the Delaunay triangulation, at initial position $x_{j0}$, with a Hookean spring. The spring constant is $k$, and the natural length is $d_{i,j}=\lvert x_{j0}-x_{i0} \rvert$, so that the initial positions are in a stationary state. The equation of motion of the $i^{\rm th}$ particle is then taken as the sum of interactions over all neighbors $j$, as 
\begin{align}
  \frac{{\rm d}\bx_i}{{\rm d}t} =
  \sum_j \left\{
    K \left(
        \left\lvert \bx_j - \bx_i \right\rvert - d_{i,j}
      \right) +
      F_{i,j}
  \right\}
  \be_{i,j}
\end{align}
where $\be_{i,j}=(\bx_j - \bx_i )/|\bx_j - \bx_i|$ denotes the unit vector in the direction of $\bx_j - \bx_i$. $K$ is equal to $k$ if $|\bx_j - \bx_i|$ is smaller than $(1+s)d_{i,j}$ and 0 if $|x_j-x_i |>(1+s)d_{i,j}$. In other words, if the distance between the neighboring particles becomes larger than $(1+s)$ times the natural length $d_{i,j}$, then the spring connecting $i$ and $j$ is cut off. $F_{i,j}$ represents the effective force generated by the surfactant. Unless both $i$ and $j$ are boundary sites, $F_{i,j}=0$. Here, the boundary sites are points where at least one connection with a neighboring point in the initial configuration has been cut off. At the boundary sites, we have
\begin{align}
F_{i,j}=\frac{F S_0}{S(t)}
\end{align}
where $S_0$ denotes the number of initial boundary sites, and $F$ defines the relative magnitude of the surfactant forces. Initially, the points within a radius of 3 from the center of the domain are removed, to form the core of the sumi-wari pattern, and we set $S_0=21$ in our simulation. $S(t)$ denotes the number of boundary sites at time $t$. The decay of $F_{i,j}$ is due to our assumption that the surfactant is adsorbed onto the boundary, and diffuses evenly along the boundary. As cracks develop, the number of boundary sites increases, and the effective force generated by the surfactant at each site weakens. If $i$ is located on the boundary and $j$ is located inside the film, then particle $i$ is attracted toward the inner region; therefore, $F_{i,j}$ serves as a kind of pressure force. If both $i$ and $j$ are located on the boundary, $i$ and $j$ are mutually attracted, and $F_{i,j}$ works as a surface tension.

\begin{figure}[tbp]
\centering
  \includegraphics[width=\linewidth]{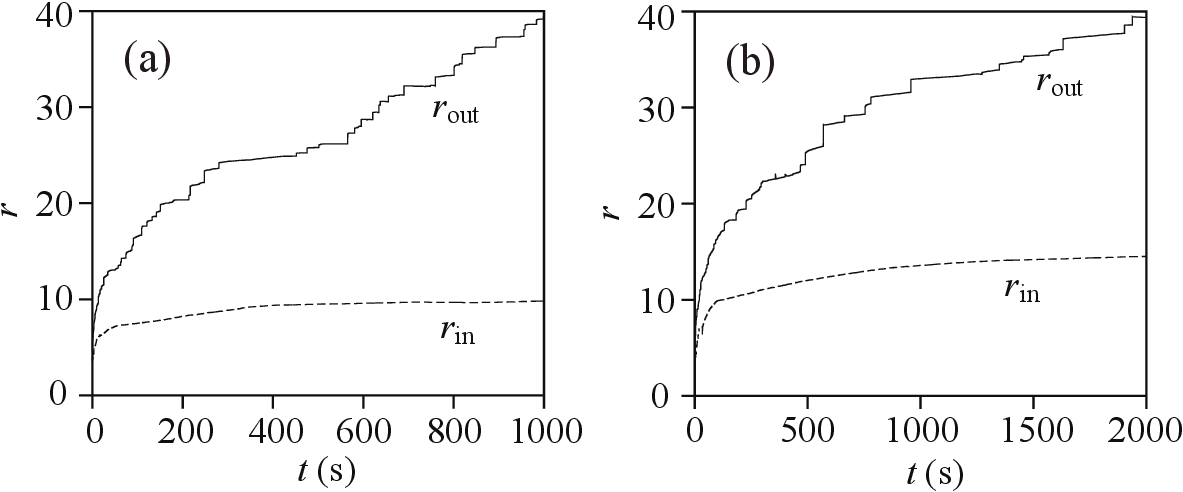}
  \caption{Time series of $r_{\rm in}$ and $r_{\rm out}$ for (a) $k=1$ and (b) $k=0.2$ for $s=0.2$, $F=4$, respectively. The solid and dashed lines represent the data of $r_{\rm out}$ and $r_{\rm in}$.}
  \label{fgr:fig8}
\end{figure}

Figure \ref{fgr:fig80} shows typical crack patterns that develop for several values of $k$, where $s=0.2$ and $F=4$. For each case, a snapshot is given with the particle positions $x_i$ plotted as dots, over the central region of $[-40,40]\times[-40,40]$. The snapshot time is different for each $k$, and the crack patterns are still growing in all cases. As shown in Fig. \ref{fgr:fig80}, the number of spikes in the sumi-wari pattern increases as $k$
decreases, and the shape of the spikes becomes more rounded. We also focus on the time series of the growth of the sumi-wari pattern in our model. Figure \ref{fgr:fig8} shows the time evolutions of the radii $r_{\rm in}$ and $r_{\rm out}$ of the inner and outer circles for (a) $k=1$ and (b) $k=0.2$. The data for Figs. \ref{fgr:fig8} (a) and (b) correspond to the cases shown in Figs. \ref{fgr:fig80} (c) and (e). When $k$ is large, the radius of the inner circle saturates at an early time, but the crack tips grow with time, making the crack propagate straight. When $k$ is small, the difference in $r_{\rm in}$ and $r_{\rm out}$ becomes small, and a round boundary with multiple protrusions appears. At moderate values of $k$, star-like crack patterns appear, which are typical of the sumi-wari pattern. 

\section{Discussion}

In this section, we compare our experimental results with those obtained from the phenomenological model and discuss the pattern formation of sumi-wari. The spring constant $k$ in the model and the subphase viscosity (i.e., glycerol concentration) in the experiments are key parameters for the formation of sumi-wari. Our measurements with AFM suggest that the average value of the stiffness, $\langle E^\ast \rangle$, of the sumi film increases as the viscosity decreases. A reasonable interpretation is that increasing $k$ in the model corresponds to an increase in the film stiffness, $E^\ast$, which can be induced by a decrease in the subphase viscosity. Moreover, the experimental results for varying viscosity show the same tendencies as the model results for the variation of $k$: (1) the number of spikes increases with $k$ in the model and decreases with the viscosity (i.e. increases with $E^\ast$) in the experiments (Figs. \ref{fgr:pattern} and  \ref{fgr:fig80}), and (2) time series of $r_{\rm in}$ and $r_{\rm out}$ for the experimental patterns have the same tendency as these obtained from the model (Figs. \ref{fgr:r-1} and \ref{fgr:fig8}). These results suggest that the sumi-wari pattern is determined by the stiffness of the film.

\section{Conclusions}

In this study, we investigated the pattern formation of sumi-wari, where the fracture dynamics of a sumi film floating on a subphase arise from differences in the interfacial tension between the cracked region of the sumi film and its surroundings. The sumi-wari pattern consists of several spikes, appearing as a sunburst or star shape. Our experiments demonstrated that the sumi-wari pattern changes with the viscosity of the subphase (i.e., glycerol amount): the number of spikes increased as the viscosity increased. Additionally, AFM measurements of the film indicated that the stiffness of the film increased as the viscosity decreased. We proposed a phenomenological model based on an overdamped equation of motion of particles coupled with breakable springs. The model included the spring constant $k$ as one of the key parameters, and representing the stiffness of the film, with large $k$ corresponding to a film with high stiffness. The model showed a similar tendency to the experimental results, including changes in pattern formation with viscosity (i.e. stiffness) and the time evolution of the characteristic pattern lengths, when we assumed that increasing $k$ in the model corresponded to decreasing viscosity in our experiments. The agreement between the experimental results and results obtained from the model suggests that the formation of the sumi-wari pattern with multiple spikes depends strongly on the stiffness of the film, which depends on the glycerol amount of the subphase under the film. The link between the experimental and numerical results demonstrates that the sumi-wari crack pattern is the physical signature of the interplay between surface tension, subphase viscosity, and film stiffness.

\section*{Author contributions}

Author contributions are defined based on the CRediT (Contributor Roles Taxonomy) and listed alphabetically.
Conceptualization: M. Shimokawa;
formal analysis: M. Shimokawa, L. Goehring, L. Pauchard, H. Sakaguchi;
funding acquisition: M. Shimokawa;
investigation: M. Shimokawa, L. Goehring, A. Kinoshita, L. Pauchard, H. Sakaguchi;
methodology: M. Shimokawa, L. Goehring, L. Pauchard, H. Sakaguchi;
project administration: M. Shimokawa;
validation: M. Shimokawa, L. Goehring, A. Kinoshita, L. Pauchard, H. Sakaguchi;
visualization: M. Shimokawa, L. Goehring, A. Kinoshita, L. Pauchard, H. Sakaguchi;
writing original draft: M. Shimokawa, L. Goehring, H. Sakaguchi;
writing review, and editing: M. Shimokawa, L. Goehring, L. Pauchard, H. Sakaguchi;

\section*{Conflicts of interest}

There are no conflicts to declare.

\section*{Data availability}

Data and codes are available upon request to the authors.

\section*{Acknowledgements}

This work was supported by Yamada Science Foundation, JSPS KAKENHI 24K06978, Sumitomo foundation Basic Science Research Projects, the JSPS Core-to-Core Program ''Advanced core-to-core network for the physics of self-organizing active matter (JPJSCCA20230002)", the Organization for the Promotion of Gender Equality at Nara Women's University, and the Exploration France 2024 at French Embassy. M.S. thanks Christophe Manquest at FAST, CNRS, Prof. Nakata and Mai Yotsumoto at Hiroshima university for technical support of experiments. M.S and A.K thank Prof. Matsuoka at Nara women's University for supporting experiments by A.K. Microscopy facilities were provided by the Imaging Suite at the School of Science and Technology at NTU. We are grateful to Dominic Eberl-Craske at Nottingham Trent University (NTU) for microscopy (AFM/SEM) support. M. S. also thanks Prof. A. Davaille at FAST, Prof. H. Kitahata at Chiba university, Prof. C. Inoue at Kyushu university, Prof. H. Uji at Nara University of Education, Prof. S. Kitsunezaki at Nara women's University for fruitful discussions.



\balance


\bibliography{ref} 
\bibliographystyle{rsc} 

\end{document}